\RequirePackage{fixltx2e}
\RequirePackage{fix-cm}

\documentclass[%
floatfix,
showkeys,
twocolumn, %
nofootinbib, %
superscriptaddress, %
]{revtex4-1}

\usepackage{cmap}

\usepackage{ucs}
\usepackage[utf8x]{inputenc}
\usepackage[T1,T2A]{fontenc}
\usepackage[german,russian,english]{babel}

\usepackage[sort&compress]{natbib}
\usepackage{amsmath}
\usepackage{amssymb}

\usepackage{mathtools}
\mathtoolsset{
showonlyrefs,
mathic = true
}

\allowdisplaybreaks

\usepackage{hyperref}
\hypersetup{backref,
 colorlinks=false}
\hypersetup{pdfborder=0 0 0}

\usepackage{breakurl}

\usepackage{microtype}
\UseMicrotypeSet[protrusion]{alltext}

\usepackage{fixltx2e}

\usepackage{nicefrac}

\usepackage{physymb}

\makeatletter
\def\ps@pprintTitle{%
     \let\@oddhead\@empty
     \let\@evenhead\@empty
     \let\@oddfoot\@empty
     \let\@evenfoot\@oddfoot}
\makeatother
\usepackage{setspace}

\renewcommand{\d}{\mathrm{d}}

\newcommand{\setR}{\mathbb{R}}

\newcommand{\setZ}{\mathbb{Z}}

\newcommand{\crd}[1]{\underline{\vphantom{j}{#1}}}
\begin{document}
\graphicspath{{image/p2p_sdu/en/}}

\title{The method of constructing models of peer to peer protocols}

\author{A. V. Demidova}
\email{avdemidova@sci.pfu.edu.ru}
\affiliation{Department of Applied Probability and Informatics\\
Peoples' Friendship University of Russia\\
Miklukho-Maklaya str. 6, Moscow, 117198, Russia}

\author{A. V. Korolkova}
\email{avkorolkova@gmail.com}
\affiliation{Department of Applied Probability and Informatics\\
Peoples' Friendship University of Russia\\
Miklukho-Maklaya str. 6, Moscow, 117198, Russia}

\author{D. S. Kulyabov}
\email{yamadharma@gmail.com}
\affiliation{Department of Applied Probability and Informatics\\
Peoples' Friendship University of Russia\\
Miklukho-Maklaya str. 6, Moscow, 117198, Russia}
\affiliation{Laboratory of Information Technologies\\
Joint Institute for Nuclear Research\\
Joliot-Curie 6, Dubna, Moscow region, 141980, Russia}

\author{L. A. Sevastyanov}
\email{leonid.sevast@gmail.com}
\affiliation{Department of Applied Probability and Informatics\\
Peoples' Friendship University of Russia\\
Miklukho-Maklaya str. 6, Moscow, 117198, Russia}
\affiliation{Bogoliubov Laboratory of Theoretical Physics\\
Joint Institute for Nuclear Research\\
Joliot-Curie 6, Dubna, Moscow region, 141980, Russia}

\thanks{Published in:
  A.~V. Demidova, A.~V. Korolkova, D.~S. Kulyabov, L.~A. Sevastianov,
  The method of constructing models of peer to peer protocols, in:
  Applied Problems in Theory of Probabilities and Mathematical
  Statistics Related to Modeling of Information Systems. The 6th
  International Congress on Ultra Modern Telecommunications and
  Control Systems. Saint-Petersburg, Russia. October 6-8, 2014, IEEE,
  2014, pp. 657--662.
}

\thanks{Sources:
  \url{https://bitbucket.org/yamadharma/articles-2013-onestep-processes}}

\begin{abstract}
  
  The models of peer to peer protocols are presented with the help of
  one-step processes. On the basis of this presentation and the method
  of randomization of one-step processes described method for
  constructing models of peer to peer protocols. As specific
  implementations of proposed method the models of FastTrack and
  Bittorrent protocols are studied.

\end{abstract}


\maketitle

  \section{Introduction}

  While constructing stochastic mathematical model there is a certain
  problem how to introduce the stochastic term which deals not with
  external impact on the system, but has a direct relationship with
  the system's structure. In to order to construct a required
  mathematical model we will consider the processes occurring in the
  system as one-step Markov processes.

  This approach allows to obtain stochastic differential equations
  with compatible stochastic and deterministic parts, since they are
  derived from the same equation. The stochastic differential
  equations theory allows qualitatively to analyse the solutions of
  these equations The Runge--Kutta methods are used to obtain the
  solutions of stochastic differential equations and for illustration
  of presented results.

  In previous studies, the authors developed a method of construction
  of mathematical model based on one-step stochastic processes, which
  describe a wide class of phenomena ~\cite{L_lit13, L_lit10::en}.
  This method presented good results for population dynamic
  models~\cite{L_lit14::en, L_lit12::en, L_lit11::en}. This method is
  also applicable to some technical problems such as p2p-networks
  simulation, in particular to the FastTrack and
  BitTorrent~\cite{kulyabov:2013:conference:mephi::en}.

  The paper proposes to use one-step stochastic processes method in
  order to construct FastTrack and BitTorrent protocol models and to
  study stochastic influence on the deterministic model.

  \section{Notations and conventions}
\label{sec:2}

  \begin{enumerate}
  \item In this paper the notation of abstract indices is
    used~\cite{penrose-rindler-1987::en}. Under the given notation, the
    tensor is denoted by an index (eg, $x^{i}$), and the
    tensor's components are denoted by an underlined index (eg,
    $x^{\crd{i}}$).

  \item Latin indices of the
    middle of the alphabet (eg, $i$, $j$, $k$) denote system space vectors. 
        Latin indices from the beginning of the
    alphabet (eg, $a$) are related to the space of Wiener's  process. Latin
    indices from the end of the alphabet (eg, $p$, $q$) are
    the indices of the Runge--Kutta methods. Greek indices (eg, $\alpha$) 
    denote a quantity of different interactions in the kinetic equations.

  \item A dot over the symbol (eg, $\dot{x}$) denotes time differentiation.

  \item A comma in the index denotes the partial derivative with respect to
    corresponding coordinate.

  \end{enumerate}

  \section{One-step processes modeling}
\label{sec:onestep}

  Under the one-step process we understand the continuous time Markov
  processes with integer state space. The transition matrix which
  allows only transitions between neighboring states. Also, these
  processes are known as birth-and-death processes.

  The state of the system is described by a state vector $x^{i} \in
  \setR^n$, where $n$~--- system dimension.

  The idea of the method is as follows. For the studied system the
  interaction scheme as a symbolic
  record of all possible interactions between system elements is introduced. The scheme
  shows the number and type of elements in certain
  interaction and the  result of the interaction. For this purpose the system state
  operators are used. The operator $n^{i \alpha}_{j} \in
  \setZ^{n}_{{}\geqslant 0} \times \setZ^{n}_{{}\geqslant 0} \times
  \setZ^{s}_{0}$ sets the state of the system before the interaction, and
  the operator $m^{i \alpha}_{j} \in \setZ^{n}_{{}\geqslant 0} \times
  \setZ^{n}_{{}\geqslant 0} \times \setZ^{s}_{0}$~--- sets the state
  after the interaction. It is also assumed that in the system 
  $s$ kinds of different interactions may occur, where $s\in \setZ_{+}$. As a
  result of the interaction , the system switches into the $x^{i}
  \rightarrow x^i + r^{i \crd{\alpha}}_{j} x^{j}$ or $x^{i}
  \rightarrow x^{i} - r^{i \crd{\alpha}}_{j} x^{j}$ states, where $r_j^{i
    \alpha} = m_j^{i \alpha} -n_j^{i \alpha}$.

  Let's introduce transition probabilities from state $x^{i}$ into
  $x^{i} + r^{i \crd{\alpha}}_{j} x^{j}$ states (the $x^{i} - r^{i
    \crd{\alpha}}_{j} x^{j}$ state). The transition probabilities of
  are assumed to be proportional to the number of possible
  interactions between elements.

  Based on the interaction schemas and transition probabilities, we
  create the master equation, decompose it into a series, leaving
  only the terms up to the second derivative. The resulting equation
  is the Fokker--Planck equation, which looks like:

\begin{equation}
  \label{eq:FP}
  \frac{\partial p}{\partial t} = -
  \partial_{i} \left[ A^{i} p \right] + \frac{1}{2} \partial_{i} \partial_{j} \left[
    B^{i j}p \right],
\end{equation}
  where
\begin{equation}
  \label{eq:kFP}
  \begin{gathered}
    A^{i} := A^{i}(x^{k}, t) = r^{i \crd{\alpha}} \left[ s^+_{\crd{\alpha}} - s^-_{\crd{\alpha}} \right], \\
    B^{i j} := B^{i j}(x^{k},t) = r^{i \crd{\alpha}} r^{j
      \crd{\alpha}} \left[ s^+_{\crd{\alpha}} - s^-_{\crd{\alpha}}
    \right].
  \end{gathered}
\end{equation}

  Here $p := p(x^{i},t)$ is a random variable $x^{i}$ density
  function, $A^{i}$ --- a drift vector, $B^{i j}$ --- a diffusion vector.

  As it is evident from \eqref{eq:kFP},  the Fokker--Planck
  equation coefficients can be obtained immediately from interaction scheme  and
  transition probabilities, i.e., for practical calculations the master
  equation is not in need.

  To get the more convenient form of the model the
  corresponding Langevin's equation is given:
\begin{equation}
  \label{eq:langevin}
  \d x^{i} = a^{i} \d t + b^i_{a} \d W^{a},
\end{equation}
  where $a^{i} := a^{i} (x^k, t)$, $b^{i}_{a} := b^{i}_{a} (x^k, t)$,
  $x^i \in \setR^n $, $W^{a} \in \mathbb{R}^m$ --- $m$-dimensional
  Wiener's process. It is implemented as $\d W = \varepsilon
  \sqrt{\d t}$, where $\varepsilon \sim N(0,1)$~--- normal distribution
  with mean equal to$0$ and variance equal to $1$.

  The relationship between the equations \eqref{eq:FP} and
  \eqref{eq:langevin}  is expressed by the following:
\begin{equation}
  \label{eq:k-langevin}
  A^{i} = a^{i}, \qquad B^{i j} = b^{i}_{a} b^{j a}.
\end{equation}

\begin{figure}%[h]
  \centering
  \includegraphics[width=\linewidth]{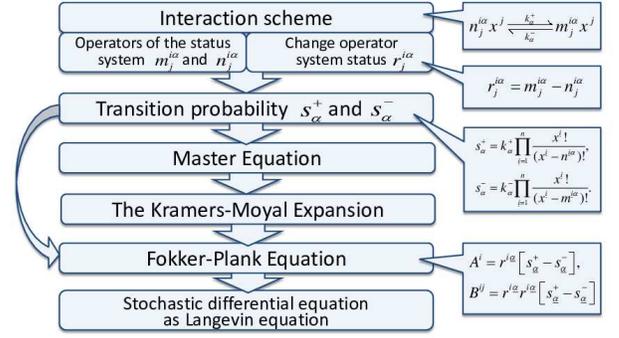}
  \caption{Method's diagram}
\label{fig:met}
\end{figure}

  Thus, for our system description the stochastic differential equation may be derived to
  from the general considerations.  This equation
  consists of two parts, one of which describes a deterministic
  behaviour of the system, and other the stochastic one. Furthermore, both sides
  of the equations are consistent, since they are derived from the same
  equation (Figure~\ref{fig:met}).

  \section{Fast Track Protocol}

  Fast Track Protocol is the p2p network protocol for Internet file sharing. 
  The file can be downloaded only from peers which possess
  the whole file.  FastTrack was originally implemented in the KaZaA
  software. It has a decentralized topology that makes it 
  very reliable. All FastTrack users are divided into two classes:
  supernodes and ordinary nodes.  The supernodes allocation is one of
  the functions of FT protocol. Supernodes are those with fast network
  connection, high-bandwidth and possibility of a fast data processing. 
  The users themselves do not know that their computer has been
  designated as a supernode.

  To upload a file, a node sends a request to the supernode, which, in
  its turn, communicates with the other nodes, etc. So, the request
  extends to a certain level protocol network which is called a
  lifetime request. After as the desired file is found, it is directly
  sent to the node bypassing the supernode from the node possessing
  the necessary file~\cite{ft1, ft2}.

\subsection{FastTrack modeling }

  Assume that the file consists of one part Thus, during one
  interaction, the node desiring to download it can
  download the entire file at once. When the download is completed, the
  node becomes supernode.

  Let $N$ denote the new node, $L$ ---  supernode and
  $\beta$ --- interaction coefficient. The new nodes appear with the intensity
  of $\lambda$, and the supernodes leave the system with the intensity of
  $\mu$.  Then the scheme of interaction and vector $\mathbf{r}$ are:
\begin{equation} 
  \label{ft:1}
  \begin{cases}
    0 \xrightarrow{\lambda } N, & r^{\crd{i}1}=(1,0) \\
    N+L \xrightarrow{\beta } 2L, & r^{\crd{i}2}=(-1,1)\\
    L \xrightarrow{\mu} 0, & r^{\crd{i}3}=(0,-1).
  \end{cases}
\end{equation}

  The first line in the diagram describes the a new
  client appearance in the system. The second line reflects the interaction of a
  new client and a seed. After this interaction a new
  seed appears.  And the third line indicates the departure of the seed from the
  system.
  Let us introduce the transition probabilities:

\begin{equation} 
  \label{ft:2}
  \begin{gathered}
    s^{+}_1 (n,l) = \lambda \\
    s^{+}_2 (n,l) = \beta nl \\
    s^{+}_3 (n,l) = \mu l.
  \end{gathered}
\end{equation}

  It is possible now to write out the Fokker-Planck equation for our
  model:
\begin{equation}
  \label{ft:3} 
  \frac{\partial p(n,l)}{\partial t} = {\partial_i}
  (A^i(n,l) p(n,l)) + \frac{1}{2} {\partial_i \partial_j} (B^{ij}(n,l)
  p(n,l)),
\end{equation}
  where the drift vector and diffusion matrix are follows:
\begin{equation}
  \begin{gathered}
    A^i: = A^i(x^k,t)= r^{i\crd{\alpha}}s^+_{\crd{\alpha}} (n,l) ,\\
    B^i:= B^{ij}(x^k,t) = r^{i\crd{\alpha}}r^{i\crd{\alpha}}
    s^+_{\crd{\alpha}} (n,l), \crd{\alpha}=1,2,3.
  \end{gathered}
\end{equation}

  At last we get:
\begin{equation} 
  \label{ft:4}
  \begin{gathered}
    \mathbf A =
    \begin{pmatrix}
      1\\
      0
    \end{pmatrix}
    \lambda +
    \begin{pmatrix}
      -1\\
      1
    \end{pmatrix}
    \beta n l +
    \begin{pmatrix}
      0\\
      -1
    \end{pmatrix}
    \mu l =
    \begin{pmatrix}
      \lambda - \beta n l\\
      \beta n l - \mu l
    \end{pmatrix}, \\
    \begin{multlined}
      \mathbf B =
      \begin{pmatrix}
        1\\
        0
      \end{pmatrix}
      (1,0) \lambda +
      \begin{pmatrix}
        -1\\
        1
      \end{pmatrix}
      (-1,1) \beta n l +
      \begin{pmatrix}
        0\\
        -1
      \end{pmatrix}
      (0,-1) \mu l = \\ =
      \begin{pmatrix}
        \lambda + \beta n l & - \beta n l \\
        - \beta n l & \beta n l + \mu l
      \end{pmatrix}.
    \end{multlined}
  \end{gathered}
\end{equation}

  The stochastic differential equation in its Langevin's form can be
  derived by using an appropriate formula.

  \subsection{Deterministic behavior}

  Since the drifts vector $A$ completely describes the deterministic
  behavior of the system, it is possible to derive an ordinary
  differential equations system, which describes the population dynamics of
  new clients and seeds.
\begin{equation}
  \label{ft:5} 
  \left \{
    \begin{aligned}
      \frac{dn}{d t}&=     \lambda - \beta n l\\
      \frac{dl}{d t}&= \beta n l - \mu l
    \end{aligned}
  \right.
\end{equation}

  \subsubsection{Steady-states}

  Let us find steady-states of the system~\eqref{ft:5} from the
  following system of equations:

\begin{equation} 
  \label{ft:6} 
  \left \{
    \begin{aligned}
      \lambda - \beta n l &=0\\
      \beta n l - \mu l &=0
    \end{aligned}
  \right.
\end{equation}

   The system~\eqref{ft:5} has the only one steady state:
\begin{equation}
(\bar{n},\bar{l})= \left ( \frac{\mu }{\beta }, \frac{\lambda }{\mu }
\right )
\end{equation}.

  To linearize the system~\eqref{ft:5}, let $n=\bar{n} + \xi $,
  $l=\bar{l} + l\eta$, where $\bar{n}$ and $\bar{l}$ are coordinates
  of stability points, $\xi$ and $\eta $ are small parameters
\begin{equation}
  \label{ft:7} 
  \left\{
    \begin{aligned}
      \frac{d\xi }{d t}&=-\beta \bar{n} \eta- \beta \bar{l} \xi \\
      \frac{d\eta }{d t}&=\beta \bar{n} \eta + \beta \bar{l} \xi - \mu
      \eta
    \end{aligned}
  \right.
\end{equation}

  In the neighborhood of the equilibrium point, the linearized 
  system is presented as following::
\begin{equation}
  \label{ft:8} 
  \left\{
    \begin{aligned}
      \frac{d\xi }{d t}&= - \mu \eta \frac{\beta \lambda }{\mu}\xi  \\
      \frac{d\eta }{d t}&= \frac{\beta \lambda }{\mu}\xi
    \end{aligned}
  \right.
\end{equation}

  Now we may find the eigenvalues of the characteristic equation:
\begin{equation}
  \label{ft:9} 
  s^2+\frac{\beta \lambda }{\mu} s + \beta
  \lambda =0.
\end{equation}

  The roots of this characteristic equation:
\begin{equation}
  \label{ft:10} 
  s_{1,2}= \frac{1}{2} \left(
    -\frac{\beta \lambda }{\mu} \pm \sqrt{ \left( \frac{\beta \lambda
        }{\mu} \right)^2 - 4 \beta \lambda} \right).
\end{equation}

  Thus, depending on the choice of parameters, the critical point has
  different types. In the case when $\beta\lambda < 4\mu^2$, the critical point represents a
  stable focus, while in the reverse case --- a steady node.  In both
  the cases, the singular point is a stable one because the real part of the
  roots of the equation is negative. Thus, depending on the choice of
  coefficient, the change of values of the variables can occur in one of two
  trajectories.

\begin{figure}%[h]
  \centering
  \includegraphics[width=\linewidth]{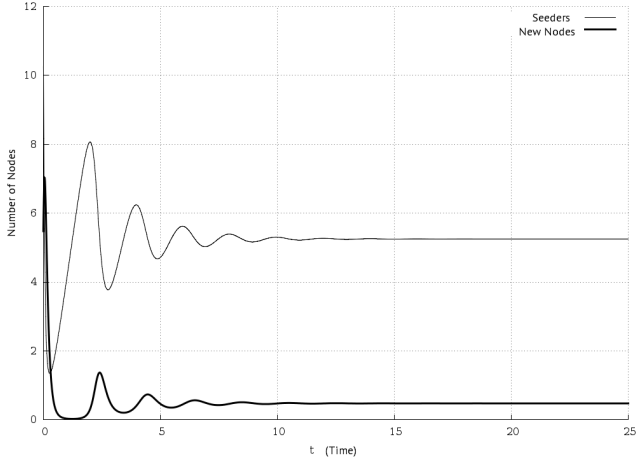}
    \caption{The time dependence of the nodes and seeds quantity 
      in the Fast Track network for the deterministic case 
      $\beta \lambda < 4\mu^2$.}
  \label{fig:ft1}
\end{figure}

\begin{figure}%[h]
  \centering
  \includegraphics[width=\linewidth]{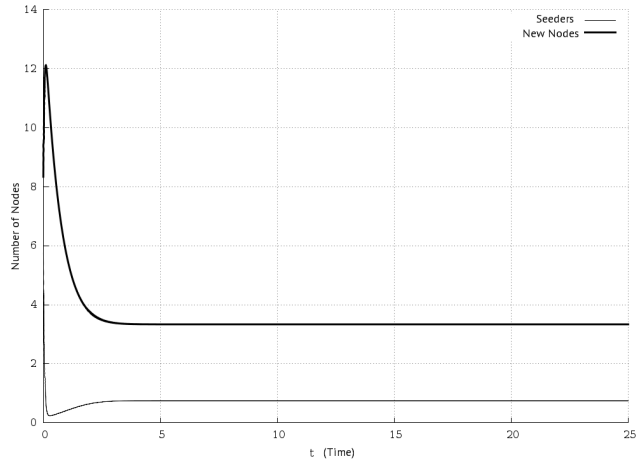}
    \caption{The time dependence of the nodes and seeds quantity  
      in Fast Track network for the deterministic case $\beta \lambda >
      4\mu^2$.}
  \label{fig:ft2}
\end{figure}

  In the case when the critical point represents a focus, the damped oscillations of the nodes
  and supernodes quantity occur ~\ref{fig:ft1}. And if the critical point
  is node, there are no oscillations in the
  trajectories~\ref{fig:ft2}. Phase portraits of the system for each
  of the two cases are plotted, respectively, in  Figs.
  ~\ref{fig:ft3} and ~\ref{fig:ft4}.

\begin{figure}%[h]
  \centering
  \includegraphics[width=\linewidth]{3}
    \caption{Phase portraits of a deterministic Fast Track system with
      various deviations  $(\Delta x, \Delta y)$  from stationary point if $\beta
      \lambda < 4\mu^2$.}
  \label{fig:ft3}
\end{figure}

\begin{figure}%[h]
  \centering
  \includegraphics[width=\linewidth]{4}
    \caption{Phase portraits of a deterministic Fast Track system with
      various deviations $(\Delta x, \Delta y)$ from stationary point if $\beta
      \lambda > 4\mu^2$.}
  \label{fig:ft4}
\end{figure}

\subsubsection{Numerical simulation of the stochastic model}

  To illustrate the obtained results the numerical modelling of
  stochastic differential equation in the Langevin's form was
  performed. The extension of Runge-Kutta methods for stochastic 
  differential equations was applied~\cite{L_lit04, L_lit01}, and a Fortran program 
  for this extension was written. The results are presented on
  Figures~\ref{fig:ft5} and~\ref{fig:all_sft}.

  Figures~\ref{fig:ft5} and~\ref{fig:all_sft} clearly  indicate that small
  stochastic terms do not substantially affect the behaviour of the
  system in the stationary point neighbourhood.  The stochastic 
  term influence exists only on the early
  evolution of the system.  After a relatively short period of time,
  the system enters the steady-state regime and differs little from
  the deterministic case.

\textbf{Conclusions}

  The obtained results indicate that the stochastic introduction in the
  steady-state regime has little effect on the behaviour of the
  system. So, the deterministic model provides the appropriate results.

  Furthermore, the proposed method allows to extend the framework of
  the tools used for the analysis, so is becomes possible to use
  ordinary stochastic differential equation (Langevin) and partial
  differential equation (Fokker-Planck) simultaneously.  Furthermore,
  as the above example indicates, in some cases a deterministic
  approach defined by the diffusion matrix is sufficient.

\begin{figure}%[h]
  \centering
  \includegraphics[width=\linewidth]{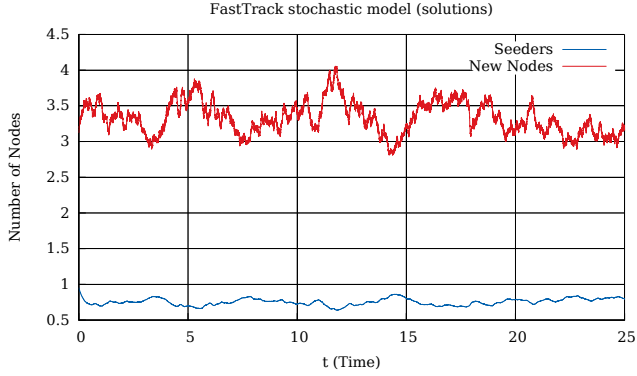}
  \caption{The time dependence of new nodes and seeds quantity 
     in the FastTrack network for the stochastic case.}
\label{fig:sft_graph}
\end{figure}

\begin{figure}%[h]
  \centering
  \includegraphics[width=\linewidth]{5}
  \caption{Phase trajectories of the stochastic FastTrack model with
    various deviations $(\Delta x, \Delta y)$ from the stationary point
    $\beta \lambda > 4\mu^2$.}
\label{fig:ft5}
\end{figure}

\begin{figure}%[]
  \centering
  \includegraphics[width=\linewidth]{all_sft}
  \caption{Phase trajectories of the stochastic FastTrack model with
    various deviations $(\Delta x, \Delta y)$ from the stationary point
    $\beta \lambda > 4\mu^2$.}
\label{fig:all_sft}
\end{figure}

\section{BitTorrent protocol}

BitTorrent is the p2p-network protocol for file sharing over the Internet. 
Files are transferred by chunks. Each torrent-client simultaneously 
downloads the needed chunks from one node and uploads available chunks to 
another node. It makes the BitTorrent protocol more flexible then the FastTrack 
one. 

\subsection{Modeling}

First, we consider a simplified model of a closed system, where the numbers 
of leechers and seeders are constant. 
Furthermore, we assume that the file consists of one chunk. 
Thus, the leecher downloads the file during only one time step and then becomes a
seeder. 

Let $N$ denote a new leecher, $C$ ---seeder, and $\beta$ --- 
interaction coefficient. Then the interaction scheme will be:

\begin{equation} 
\label{bt:1}
N+C \xrightarrow{\beta } 2C, \qquad r^{\crd{i}2}=(-1,1).
\end{equation}

The scheme reflects that after the leecher interaction with the seeder, the 
leecher disappears and another seeder appears. 

Next, let $n$ be the number of new nodes, and $c$ ---
the number of seeders in the system.

The transition probabilities:
\begin{equation} 
\label{bt:2}
s^{+} (n,c) = \beta nc.
\end{equation}

The Fokker--Planck's equation for this model:
\begin{equation} 
\label{bt:3}
\frac{\partial p(n,c)}{\partial t} = 
{\partial_i} (A^i(n,c) p(n,c))
+\frac{1}{2} {\partial_i \partial_j} (B^{ij}(n,c) p(n,c)),
\end{equation}
where the drift vector and the diffusion matrix are presented as following:
\begin{equation}
  \begin{gathered}
     A^i(n,c)= r^{i\crd{\alpha}}s^+_{\crd{\alpha}} (n,l) ,\\
     B^i(n,c) = r^{i\crd{\alpha}}r^{i\crd{\alpha}} s^+_{\crd{\alpha}} (n,l).
  \end{gathered}
\end{equation}

Thus, we obtain:
\begin{equation} 
\label{bt:4}
  \begin{gathered}
    \mathbf A =
    \begin{pmatrix}
      -1\\
      1
    \end{pmatrix}
    \beta n c +
     =
    \begin{pmatrix}
      - \beta n l\\
      \beta n l 
    \end{pmatrix}, \\
    \begin{multlined}
      \mathbf B =
       \begin{pmatrix}
        -1\\
        1
      \end{pmatrix}
      (-1,1) \beta n c  = 
      \begin{pmatrix}
         \beta n c & - \beta n c \\
        - \beta n c& \beta n c
      \end{pmatrix}.
    \end{multlined}
  \end{gathered}
\end{equation}

The stochastic differential equation in the Langevin's form can be 
obtained with the help of the appropriate formula.

It is also possible to write out differential equations which describe the
deterministic behaviour of the system:
\begin{equation} 
\label{bt:5}
\left \{
\begin{aligned}
 \frac{dn}{d t}&=      - \beta n c\\
 \frac{dc}{d t}&=     \beta n c 
\end{aligned}
\right.
\end{equation}

Next, we consider the open system in which new clients appear with the intensity $\lambda $, 
and seeders leave it with  the intensity $\mu $. Now, the scheme of interaction has the form of:

\begin{equation} 
\label{bt:6}
\begin{aligned}
0 \xrightarrow{\lambda } N, & r^{\crd{i}1}=(1,0),\\
N+C \xrightarrow{\beta } 2C, & r^{\crd{i}2}=(-1,1),\\
C  \xrightarrow{\mu } 0, & r^{\crd{i}3}=(0,-1).
\end{aligned}
\end{equation}

The first line in the scheme describes the appearance of the new peer in 
the system, the second line describes the interactions between the new 
peer and the seeder, after which a new seeder appears. And the third 
line meaning is that the seeder leaves the system. 

Let $n$ denote the number of new clients and 
$c$ --- the number of seeders in the system.

This system is equivalent to the Fast Track model up to notation. 

Now consider a system in which downloaded files consist of $m$ chunks. 
The system consists of: 

\begin{itemize}
\item Peers ($N$) are the clients without any chunk of the
  file.
\item Leechers ($L$) are the clients who have already downloaded a
  number of chunks of the file and can share them with new peers or
  other leechers.
\item Seeders ($C$) are the clients who have the whole file and they
  only can share the file.
\end{itemize}

In addition, $n$ is the number of new peers, and $c$ --- number of 
seeders in the system, $l_i$ --- number of leechers with exactly 
$i$ chunks of the file, where $i = \overline{i, n-1}$. Also, let $\bar 
{L}_i$ be the number of leechers with any chunk of the file of interest for
leecher $ L_i $ and the $ \bar{l}_i $ is  their amount. 

For this scheme it is possible to write out the following types of relations:
\begin{equation} 
\label{bt:7}
\begin{aligned}
0  \xrightarrow{\lambda } & N, \\
N+C  \xrightarrow{\beta } & L_1+C, \\
N+L_i  \xrightarrow{\beta_i } & L_1+L_i, \\
L_i + \bar{L}_i  \xrightarrow{\delta_i } & L_{i+1}+\bar{L}_i, \\
L_i + C  \xrightarrow{\gamma_i } & L_{i+1}+C, \\
L_{m-1} + \bar{L}_{m-1}  \xrightarrow{\gamma_{m-1} } & C+\bar{L}_{m-1}, \\
L_{m-1} + C  \xrightarrow{\gamma } & 2C, \\
C  \xrightarrow{\mu } & 0. 
\end{aligned}
\end{equation}

  On every interaction step one chunk of file is transferred from one
  peer to another. The first relation describes the appearance of a
  new peer in a system with the intensity $\lambda$.

  The second and third relations describe the interaction of a new
  peer with a seeder or a leecher with the interaction coefficients $\beta$
  and $\beta i$, $(i=\overline{i, m-1})$. As the result of interaction, the
  peer transforms into a leecher from the $L_1$ class.  The fourth and fifth
  relations describe the leecher $L_i$ interaction with the seeder and other
  leechers with the coefficients $\delta_i$ and $\gamma_i$ $(i=\overline{i,
    m-2})$.  As the result of this interaction, the leecher gets one chunk
  of a file and becomes the $L_{i+1}$-class leacher.  The sixth and seventh
  relations describe the transformation of leecher into seeders with
  the coefficients $\gamma_{m-1} $ and $\gamma$ (the leecher downloads the last
  file chunk). The last relation describes the seeder departure from the
  system with the intensity $\mu$.

  The vectors $r^{i\crd{\alpha}}=(n,l_1,l_2,...,l_{m-1},c)$
  and transition probabilities $s^+_{\crd{\alpha}}$:
\begin{equation} 
\label{bt:8}
\begin{gathered}
r^{1}  =(1,0,0,...,0), \\
r^{2}  =r_i^3=(-1,1,0,...,0), i=\overline{i, m-1} \\
r_i^4  =r_i^5=(0,...,-1,1,...,0), i=\overline{i, m-2} \\
r^{6}  =r^7=(0,0,...,-1,1), \\
r^{8}  =(0,0,...,-1).
\end{gathered}
\end{equation}
\begin{equation} 
\label{bt:9}
\begin{gathered}
s^{+}_1  =\lambda, \\
s^{+}_2  =\beta n c, \\
s^{+}_{3i}  =\beta_i n l_i, \\
s^{+}_{4i}  =\delta_i l_i \bar{l}_i, i=\overline{i, m-1}\\
s^{+}_{5i}  =\gamma_i l_i c, i=\overline{i, m-2}\\
s^{+}_{6}  =\gamma_{m-1} l_{m-1} \bar{l}_{m-1}, \\
s^{+}_{7}  =\gamma l_{m-1} c, \\
s^{+}_{8}  =\mu c. \\
\end{gathered}
\end{equation}

  For this model, which is similar to the previous one, we can write out the
  Fokker-Planck's equation. But for deterministic behaviour description,
  it's  enough to write out the matrix $A$.
\begin{equation}
\label{bt:10}
    \mathbf A =
    \begin{pmatrix}
    \lambda - \beta n c - \sum_{i=1}^{m-1} \beta_i n l_i \\
    \beta n c + \sum_{i=1}^{m-1} \beta_i n l_i -\delta_1 l_1 \bar{l}_1 - \gamma_1 l_1 c \\
      \delta_1 l_1 \bar{l}_1 + \gamma_1 l_1 c -  \delta_2 l_2 \bar{l}_2 - \gamma_2 l_2 c  \\
     \ldots \\
     \begin{multlined}
       \delta_{m-2} l_{m-2} \bar{l}_{m-2} + \gamma_{m-2} l_{m-2} c -
       {} \\ {} -
       \delta_{m-1} l_{m-1} \bar{l}_{m-1} - \gamma_{m-1} l_{m-1} c 
     \end{multlined}
     \\ 
      \delta_{m-1} l_{m-1} \bar{l}_{m-1} + \gamma_{m-1} l_{m-1} c  - \mu c 
    \end{pmatrix}.
\end{equation}

As a result, we obtain a system of differential equations
describing the dynamics of new peers, leechers and seeders:

\begin{equation} \label{bt:11}
\left \{
\begin{gathered}
  \frac{d n}{d t} =      \lambda - \beta n c - \sum_{i=1}^{m-1} \beta_i n l_i, \\
 \frac{d l_1}{d t}=     \beta n c + \sum_{i=1}^{m-1} \beta_i n l_i -\delta_1 l_1 \bar{l}_1 - \gamma_1 l_1 c, \\ 
 \frac{d l_2}{d t}=  \delta_1 l_1 \bar{l}_1 + \gamma_1 l_1 c -  \delta_2 l_2 \bar{l}_2 - \gamma_2 l_2 c,  \\
    \ldots \\
\begin{multlined}
 \frac{d l_{m-1}}{d t}=  \delta_{m-2} l_{m-2} \bar{l}_{m-2} + 
 \gamma_{m-2} l_{m-2} c - {} \\ {} - 
 \delta_{m-1} l_{m-1} \bar{l}_{m-1} -
 \gamma_{m-1} l_{m-1} c,
\end{multlined}
\\ 
\frac{d c}{d t}= \delta_{m-1} l_{m-1} \bar{l}_{m-1} + \gamma_{m-1} l_{m-1} c  - \mu c. 
\end{gathered}
\right.
\end{equation}

 Let's suppose that 
  $\delta=\delta_{1}=\delta_{2}=...=\delta_{m-1}=const$, then let's  sum the
  equations in our system from the second one to the $m+1$-th. If we denote
  leechers and seeders as $l = l_1 + l_2 + ... + l_{m-1} + c$  the
  system of the equations may be simplified as follows:

\begin{equation} 
\label{bt:12}
\left \{
\begin{aligned}
 \frac{d n}{d t}&=      \lambda - \beta n (l+c), \\
 \frac{d (l+c)}{d t}&=     \beta n (l+c)  - \mu c. 
\end{aligned}
\right.
\end{equation}

\section{Conclusion}

\begin{enumerate}
\item In this paper the method of stochastic models construction by use of one-step stochastic 
processes is described.  The proposed method provides an universal algorithm of 
deriving stochastic differential equations for such systems. It's also shown that there are 
two way of stochastic system's description: with the help of 
partial differential equation (Fokker-Plank) and ordinary  differential 
equations (Langevin). 

\item In order to study influence of the stochastic term of an equation the Fast Track and Bittorrent
protocol models were discussed. The results of this study 
indicate, that near the stationary points the stochastic influence is minimal, and 
that's because the deterministic model gives very good results. In addition, 
as it was shown by the above example, in some cases, in order to examine the system it 
is enough to study its deterministic approximation, which is described 
by the drift matrix. 
\end{enumerate}

\bibliographystyle{abbrvnat}

\bibliography{bib/p2p_sdu/p2p_sdu}

\end{document}

\graphicspath{{image/p2p_sdu/ru/}}

\title{Методика построения моделей пиринговых протоколов}

\author{А. В. Демидова}
\email{avdemidova@sci.pfu.edu.ru}
\affiliation{Кафедра прикладной информатики и теории вероятностей,\\
  Российский университет дружбы народов,\\
  ул. Миклухо-Маклая, д.6, Москва, Россия, 117198}

\author{А. В. Королькова}
\email{avkorolkova@gmail.com}
\affiliation{Кафедра прикладной информатики и теории вероятностей,\\
  Российский университет дружбы народов,\\
  ул. Миклухо-Маклая, д.6, Москва, Россия, 117198}

\author{Д. С. Кулябов}
\email{yamadharma@gmail.com}
\affiliation{Кафедра прикладной информатики и теории вероятностей,\\
  Российский университет дружбы народов,\\
  ул. Миклухо-Маклая, д.6, Москва, Россия, 117198}
\affiliation{Лаборатория информационных технологий,\\
Объединённый институт ядерных исследований,\\
ул. Жолио-Кюри 6, Дубна, Московская область, Россия, 141980}

\author{Л. А. Севастьянов}
\email{leonid.sevast@gmail.com}
\affiliation{Кафедра прикладной информатики и теории вероятностей,\\
  Российский университет дружбы народов,\\
  ул. Миклухо-Маклая, д.6, Москва, Россия, 117198}
\affiliation{Лаборатория теоретической физики,\\
  Объединённый институт ядерных исследований,\\
  ул. Жолио-Кюри 6, Дубна, Московская область, Россия, 141980}

  \thanks{Опубликовано в: 
    \emph{Demidova~A.~V., Korolkova~A.~V., Kulyabov~D.~S.,
      Sevastianov~L.~A.}  The method of constructing models of peer to
    peer protocols~// Applied Problems in Theory of Probabilities and
    Mathematical Statistics Related to Modeling of Information
    Systems. The 6th International Congress on Ultra Modern
    Telecommunications and Control Systems. Saint-Petersburg, Russia.
    October 6-8, 2014.~--- IEEE, 2014.~--- P.~657--662.  }

\thanks{Исходные тексты:
  \url{https://bitbucket.org/yamadharma/articles-2013-onestep-processes}}

\begin{abstract}
  Пиринговые протоколы представлены как одношаговые процессы. На
  основе этого представления и метода стохастизации одношаговых
  процессов описывается методика построения моделей пиринговых
  протоколов. В качестве конкретных реализаций приводятся модели
  протоколов Fasttrack и базового Bittorrent-подобного протокола.

\end{abstract}

  \keywords{стохастические дифференциальные уравнения; основное
    кинетическое уравнения; уравнение Фоккера--Планка; FastTrack;
    BitTorrent}

\maketitle

\section{Введение}

  При стохастизации математических моделей возникает проблема, как
  ввести стохастический член, который интерпретируется не как внешнее
  случайное воздействие на систему, а имеет непосредственную связь с
  ее структурой. Для получения стохастических моделей предлагается
  рассматривать процессы, происходящие в системе, как одношаговые
  марковские процессы. Такой подход позволяет получать стохастические
  дифференциальные уравнения с согласованными стохастической и
  детерминистической частями, так как они выводятся из одно и того же
  уравнения. Привлечение теории стохастических дифференциальных
  уравнений позволяет провести качественный и численный анализ
  поведения решений уравнений для полученной стохастической модели.
  Для иллюстрации результатов предлагается использовать численные
  Рунге--Кутты разных порядков построения решений стохастических
  дифференциальных уравнений.

  В предыдущих работах авторов разработан метод построения одношаговых
  стохастических моделей, который позволяет моделировать широкий класс
  явлений~\cite{L_lit13, L_lit10}. Данный метод показал хорошие
  результаты для популяционной динамики~\cite{L_lit14, L_lit12,
    L_lit11} . Его также можно применить к техническим задачам таким
  как peer-to-peer сети, в частности к моделированию протокола
  FastTrack и BitTorrent~\cite{kulyabov:2013:conference:mephi}.

  В работе предлагается применение данного метода для построения
  моделей протоколов FastTrack и BitTorrent и изучение влияния
  введения стохастики в детерминистическую модель.

\section{Обозначения и соглашения}
\label{sec:2}

  \begin{enumerate}
  \item В работе используется нотация абстрактных
    индексов~\cite{penrose-rindler-1987}. В данной нотации тензор как
    целостный объект обозначается просто индексом (например, $x^{i}$),
    компоненты обозначаются подчёркнутым индексом (например,
    $x^{\crd{i}}$).

  \item Будем придерживаться следующих соглашений. Латинские индексы
    из середины алфавита ($i$, $j$, $k$) будут относиться к
    пространству векторов состояний системы. Латинские индексы из
    начала алфавита ($a$) будут относиться к пространству винеровского
    процесса. Латинские индексы из конца алфавита ($p$, $q$) будут
    относиться к индексам метода Рунге--Кутты. Греческие индексы
    ($\alpha$) будут задавать количество разных взаимодействий в
    кинетических уравнениях.

  \item Точкой над символом обозначается дифференцирование по времени.

  \item Запятой в индексе обозначается частная производная по
    соответствующей координате.

  \end{enumerate}

\section{Моделирование одношаговых процессов}
\label{sec:onestep}

  Под одношаговыми процессами мы будем понимать марковские процессы с
  непрерывным временем, принимающие значения в области целых чисел,
  матрица перехода которых допускает только переходы между соседними
  участками. Также эти процессы известны под названиями процессов
  рождения--гибели.

  Состояние системы будем описывать вектором состояния $x^{i} \in
  \setR^n$, где $n$~--- размерность системы

  Идея метода состоит в следующем. Для исследуемой системы, состояние
  которой будем описывать вектором состояния $x^{i} \in \setR^n$, где
  $n$~--- размерность системы, можно записать схему
  взаимодействия. Т.е. символическую запись всех возможных
  взаимодействий между элементами системы, которая показывает сколько
  и каких элементов во взаимодействие какого типа вступают и что
  получилось в результате. Для этого используются операторы состояния
  системы. Оператор $n^{i \alpha}_{j} \in \setZ^{n}_{{}\geqslant 0}
  \times \setZ^{n}_{{}\geqslant 0} \times \setZ^{s}_{0}$ задаёт
  состояние системы до взаимодействия, оператор $m^{i \alpha}_{j} \in
  \setZ^{n}_{{}\geqslant 0} \times \setZ^{n}_{{}\geqslant 0} \times
  \setZ^{s}_{0}$~--- после. Также считается, что в системе может
  происходить $s$ видов различных взаимодействий, где $s\in
  \setZ_{+}$. И в результате взаимодействия система переходит в
  состояние $x^{i} \rightarrow x^i + r^{i \crd{\alpha}}_{j} x^{j}$ или
  $x^{i} \rightarrow x^{i} - r^{i \crd{\alpha}}_{j} x^{j}$, где
  $r_j^{i \alpha} = m_j^{i \alpha} -n_j^{i \alpha}$~---оператор
  изменения состояния системы.

  Далее предлагается записать вероятности переходов из состояния
  $x^{i}$ в состояние $x^{i} + r^{i \crd{\alpha}}_{j} x^{j}$ (в
  состояние $x^{i} - r^{i \crd{\alpha}}_{j} x^{j}$), которые
  предполагаются пропорциональными числу возможных взаимодействий
  между элементами.

  % и определяются выражениями:
  % \begin{equation}
  %   \label{eq:s-pm}
  %   \begin{gathered}
  %     s^+_{\alpha} = k^{+}_{\alpha} \prod_{\crd{i}}
  %     \frac{x^{\crd{i}}!}{(x^{\crd{i}} - n^{\crd{i} \alpha})!},       \\
  %     s^-_{\alpha} = k^{-}_{\alpha} \prod_{\crd{i}}
  %     \frac{x^{\crd{i}}!}{(x^{\crd{i}}-m^{\crd{i} \alpha})!}.
  %   \end{gathered}
  % \end{equation}

  На основании схем взаимодействия и вероятностей переходов мы строим
  основное кинетическое уравнение, раскладываем его в ряд, оставляя
  только члены до второй производной включительно. Получившееся
  уравнение будет уравнением Фоккера--Планка, которое будет иметь вид:

\begin{equation}
  \label{eq:FP}
  \frac{\partial p}{\partial t} = -
  \partial_{i} \left[ A^{i} p \right] + \frac{1}{2} \partial_{i} \partial_{j} \left[
    B^{i j}p \right],
\end{equation}
  где
\begin{equation}
  \label{eq:kFP}
  \begin{gathered}
    A^{i} := A^{i}(x^{k}, t) = r^{i \crd{\alpha}} \left[ s^+_{\crd{\alpha}} - s^-_{\crd{\alpha}} \right], \\
    B^{i j} := B^{i j}(x^{k},t) = r^{i \crd{\alpha}} r^{j
      \crd{\alpha}} \left[ s^+_{\crd{\alpha}} - s^-_{\crd{\alpha}}
    \right].
  \end{gathered}
\end{equation}

  Здесь $p := p(x^{i},t)$ и имеет смысл плотности распределения
  случайной величины $x^{i}$, $A^{i}$ --- вектор сноса, $B^{i j}$ ---
  вектор диффузии.

  Как видно из \eqref{eq:kFP}, коэффициенты уравнения Фоккера--Планка
  можно получить сразу после записи схемы взаимодействия и
  вероятностей перехода, то есть в практических расчётах записывать
  основное кинетическое уравнение нет необходимости.

  Для получения более привычного вида модели записываем
  соответствующее ему уравнение Ланжевена:
\begin{equation}
  \label{eq:langevin}
  \d x^{i} = a^{i} \d t + b^i_{a} \d W^{a},
\end{equation}
  где $a^{i} := a^{i} (x^k, t)$, $b^{i}_{a} := b^{i}_{a} (x^k, t)$,
  $x^i \in \setR^n $ --- вектор состояния системы, $W^{a} \in
  \mathbb{R}^m$ --- $m$-мерный винеровский процесс. Винеровский
  процесс реализуется как $\d W = \varepsilon \sqrt{\d t}$, где
  $\varepsilon \sim N(0,1)$~--- нормальное распределение со средним
  $0$ и дисперсией $1$. Латинскими индексами из середины алфавита
  обозначаются величины, относящиеся к векторам состояний (размерность
  пространства $n$), а латинскими индексами из начала алфавита
  обозначаются величины, относящиеся к вектору винеровского процесса
  (размерность пространства $m \leqslant n$).

  При этом связь между уравнениями \eqref{eq:FP} и \eqref{eq:langevin}
  выражается следующими соотношениями:
\begin{equation}
  \label{eq:k-langevin}
  A^{i} = a^{i}, \qquad B^{i j} = b^{i}_{a} b^{j a}.
\end{equation}

  Таким образом для описания системы из общих соображений можно
  получить стохастическое дифференциальное уравнение. Это уравнение
  состоит из двух частей, один из которых описывает детерминистическое
  поведение системы, а другой стохастическое. Кроме того, обе части
  уравнения являются согласованными, т. к. получены из одного и того
  же уравнения (схема на рис.~\ref{fig:met}).

\begin{figure}%[h]
  \centering
  \includegraphics[width=\linewidth]{met}
  \caption{Схема метода}
\label{fig:met}
\end{figure}

\section{Протокол Fast Track}

  Fast Track --- одноранговый (P2P) сетевой протокол для
  кооперативного обмена файлами через Интернет. Закачка данных
  осуществляется только из источников, содержащих полные
  файлы. FastTrack первоначально был реализован в программе
  KaZaA. Сеть, основанная на работе прокола FastTrack, имеет
  децентрализованную топологию, что делает ее работу очень надежной. В
  сети пользователи разделены на два класса: суперузлы и простые узлы
  (supernodes и ordinary nodes). Выделение суперузлов является одной
  из функций протокола и на эту роль выбираются узлы с быстрым
  подключением к сети, высокой пропускной способностью и возможностью
  быстрой обработки данных. При этом владельцы компьютеров не знают,
  что их компьютер был назначен в качестве суперузла.

  Для того, чтобы загрузить файл, узел посылает запрос суперузлу,
  который в свою очередь взаимодействует с другими узлами и т.д. Таким
  образом запрос распространяется до определенного протоколом уровня
  сети и называется временем жизни запроса (Time to live). После того,
  как нужный файл будет найден, он передается непосредственно узлу,
  его запросившему, от узла, который имеет этот файл, минуя
  суперузел~\cite{ft1, ft2}.

  \subsection{Моделирование}

  Сделаем предположение, что файл состоит из одной части. Таким
  образом за один шаг взаимодействия нового узла, желающего скачать
  файл, и узла, раздающего файл, новый узел скачивает весь файл и
  становится раздающим узлом.

  Пусть $N$ --- это обозначение нового узла, $L$ --- это раздающий
  узел, а $\beta $ --- коэффициент взаимодействия. Новые узлы могут
  приходить в систему с интенсивностью $\lambda $, а раздающие узлы
  уходить из нее с интенсивностью $\mu$. Тогда схема взаимодействия и
  вектор $\mathbf r$ будет иметь вид:
\begin{equation} 
  \label{ft:1}
  \begin{cases}
    0 \xrightarrow{\lambda } N, & r^{\crd{i}1}=(1,0) \\
    N+L \xrightarrow{\beta } 2L, & r^{\crd{i}2}=(-1,1)\\
    L \xrightarrow{\mu} 0, & r^{\crd{i}3}=(0,-1).
  \end{cases}
\end{equation}

  Первая строка в схеме описывает появление нового клиента в
  системе. Вторая строка отражает взаимодействие нового клиента и
  сида, в результате которого появляется новый сид. А третья – это
  уход сида из системы.
  Запишем вероятности переходов:

\begin{equation} 
  \label{ft:2}
  \begin{gathered}
    s^{+}_1 (n,l) = \lambda \\
    s^{+}_2 (n,l) = \beta nl \\
    s^{+}_3 (n,l) = \mu l.
  \end{gathered}
\end{equation}

  Далее можно записать уравнение Фоккера-Планка для данной модели:
\begin{equation}
  \label{ft:3} 
  \frac{\partial p(n,l)}{\partial t} = {\partial_i}
  (A^i(n,l) p(n,l)) + \frac{1}{2} {\partial_i \partial_j} (B^{ij}(n,l)
  p(n,l)),
\end{equation}
  где вектор сносов и матрица диффузии имеют следующий вид:
\begin{equation}
  \begin{gathered}
    A^i: = A^i(x^k,t)= r^{i\crd{\alpha}}s^+_{\crd{\alpha}} (n,l) ,\\
    B^i:= B^{ij}(x^k,t) = r^{i\crd{\alpha}}r^{i\crd{\alpha}}
    s^+_{\crd{\alpha}} (n,l), \crd{\alpha}=1,2,3.
  \end{gathered}
\end{equation}

  Таким образом получаем:
\begin{equation} 
  \label{ft:4}
  \begin{gathered}
    \mathbf A =
    \begin{pmatrix}
      1\\
      0
    \end{pmatrix}
    \lambda +
    \begin{pmatrix}
      -1\\
      1
    \end{pmatrix}
    \beta n l +
    \begin{pmatrix}
      0\\
      -1
    \end{pmatrix}
    \mu l =
    \begin{pmatrix}
      \lambda - \beta n l\\
      \beta n l - \mu l
    \end{pmatrix}, \\
    \begin{multlined}
      \mathbf B =
      \begin{pmatrix}
        1\\
        0
      \end{pmatrix}
      (1,0) \lambda +
      \begin{pmatrix}
        -1\\
        1
      \end{pmatrix}
      (-1,1) \beta n l +
      \begin{pmatrix}
        0\\
        -1
      \end{pmatrix}
      (0,-1) \mu l = \\ =
      \begin{pmatrix}
        \lambda + \beta n l & - \beta n l \\
        - \beta n l & \beta n l + \mu l
      \end{pmatrix}.
    \end{multlined}
  \end{gathered}
\end{equation}

  Стохастическое дифференциальное уравнение в форме Ланжевена можно
  получить воспользовавшись соответствующей формулой.

  \subsection{Детерминистическое поведение}

  Так как вектор сносов $A$ полностью описывает детерминистическое
  поведение системы можно получить систему обыкновенных
  дифференциальных уравнений, описывающих динамику численности новых
  клиентов и сидов:
\begin{equation}
  \label{ft:5} 
  \left \{
    \begin{aligned}
      \frac{dn}{d t}&=     \lambda - \beta n l\\
      \frac{dl}{d t}&= \beta n l - \mu l
    \end{aligned}
  \right.
\end{equation}

  \subsubsection{Стационарные состояния}

  Найдём стационарные состояния системы~\eqref{ft:5}, которые
  являются решением системы уравнений:

\begin{equation} 
  \label{ft:6} 
  \left \{
    \begin{aligned}
      \lambda - \beta n l &=0\\
      \beta n l - \mu l &=0
    \end{aligned}
  \right.
\end{equation}

  Система~\eqref{ft:5} имеет одностационарное состояние:
\begin{equation}
(\bar{n},\bar{l})= \left ( \frac{\mu }{\beta }, \frac{\lambda }{\mu }
\right )
\end{equation}.

  \subsubsection{Исследование линеаризованной устойчивости}
  \subsubsection{Study of linearized stability}

  Линеаризуем систему~\eqref{ft:5}. Пусть $n=\bar{n} + \xi $,
  $l=\bar{l} + l\eta$, где $\bar{n}$ и $\bar{l}$ --- координаты точки
  равновесия, а $\xi $ и $\eta $ --- малые возмущения:
\begin{equation}
  \label{ft:7} 
  \left\{
    \begin{aligned}
      \frac{d\xi }{d t}&=-\beta \bar{n} \eta- \beta \bar{l} \xi \\
      \frac{d\eta }{d t}&=\beta \bar{n} \eta + \beta \bar{l} \xi - \mu
      \eta
    \end{aligned}
  \right.
\end{equation}

  Запишем линеаризованную систему в окрестности точки равновесия:
\begin{equation}
  \label{ft:8} 
  \left\{
    \begin{aligned}
      \frac{d\xi }{d t}&= - \mu \eta \frac{\beta \lambda }{\mu}\xi  \\
      \frac{d\eta }{d t}&= \frac{\beta \lambda }{\mu}\xi
    \end{aligned}
  \right.
\end{equation}

  Найдём собственные значения характеристического уравнения, которое
  имеет вид:
\begin{equation}
  \label{ft:9} 
  s^2+\frac{\beta \lambda }{\mu} s + \beta
  \lambda =0.
\end{equation}

  Далее запишем корни этого характеристического уравнения:
\begin{equation}
  \label{ft:10} 
  s_{1,2}= \frac{1}{2} \left(
    -\frac{\beta \lambda }{\mu} \pm \sqrt{ \left( \frac{\beta \lambda
        }{\mu} \right)^2 - 4 \beta \lambda} \right).
\end{equation}

  Таким образом, в зависимости от выбора параметров особая точка может
  иметь разный характер. Так при $\beta \lambda < 4\mu^2$ особая точка
  является устойчивым фокусом, а при обратном соотношении ---
  устойчивый узел. В обоих случаях особая точка является устойчивой,
  так как действительная часть корней уравнения отрицательная. Таким
  образом, в зависимости от выбора значений коэффициентов, изменения
  переменных системы может происходить по одной из двух
  траекторий. Если особая точка является фокусом, то в системе
  происходят затухающие колебания численностей новых и раздающих
  узлов~\ref{fig:ft1}. А в узловом случае приближение численностей к
  стационарным значениям происходит в бесколебательном
  режиме~\ref{fig:ft2}. Фазовые портреты системы для каждого из двух
  случаев изображены, соответственно, на графиках ~\ref{fig:ft3}
  и~\ref{fig:ft4}.

\begin{figure}%[h]
  \centering
  \includegraphics[width=\linewidth]{1}
    \caption{Зависимость числа новых и раздающих узлов от времени в
      сети Fast Track для детерминистического случая при 
      $\beta \lambda < 4\mu^2$.}
  \label{fig:ft1}
\end{figure}

\begin{figure}%[h]
  \centering
  \includegraphics[width=\linewidth]{2}
    \caption{Зависимость числа новых и раздающих узлов от времени в
      сети Fast Track для детерминистического случая при $\beta
      \lambda > 4\mu^2$.}
  \label{fig:ft2}
\end{figure}

\begin{figure}%[h]
  \centering
  \includegraphics[width=\linewidth]{3}
    \caption{Фазовые портреты детерминистической системы Fast Track с
      различными отклонениями $(\Delta x, \Delta y)$ от стационарной
      точки при $\beta \lambda < 4\mu^2$.}
  \label{fig:ft3}
\end{figure}

\begin{figure}%[h]
  \centering
  \includegraphics[width=\linewidth]{4}
    \caption{ Фазовые портреты детерминистической системы Fast Track с
      различными отклонениями $(\Delta x, \Delta y)$ от стационарной
      точки при $\beta \lambda > 4\mu^2$.}
  \label{fig:ft4}
\end{figure}

  \subsubsection{Численное моделирование стохастической модели}

  Для иллюстрации полученных результатов было проведено численное
  моделирование стохастического дифференциального уравнения в форме
  Ланжевена. Для численного решения стохастических дифференциальных
  уравнений использован метод, заключающийся в распространении методов
  Рунге-Кутты на случай стохастических дифференциальных
  уравнений~\cite{L_lit04, L_lit01}, реализованный на языке
  Фортран. Результаты численного моделирования приведены на
  графиках~\ref{fig:ft5} и~\ref{fig:all_sft}.

  На рисунках~\ref{fig:ft5} и~\ref{fig:all_sft} наглядно видно, что
  введение малых стохастических членов существенно не влияет на
  поведение системы в близи узловой точки при большом числе раздающих
  узлов. Последствия введения стохастики ощущаются лишь в начале
  эволюции системы. По прошествии сравнительно небольшого отрезка
  времени система входит в стационарный режим и мало отличается от
  детерминированного случая.

\subsubsection{Выводы}

  Полученные результаты показывают, что введение стохастики в
  стационарном режиме слабо влияет на поведение системы, поэтому при
  ее изучении можно рассматривать детерминистическую модель. Кроме
  того, предложенный метод позволяет расширить аппарат инструментов,
  используемых для анализа модели, так как одновременно при применении
  данного подхода для описания системы можно получить обыкновенное
  стохастическое дифференциальное уравнение и уравнение в частных
  производных в форме уравнения Фоккера-Планка. Кроме того, как
  показал рассмотренный пример в некоторых случаях для изучения
  системы можно рассматривать ее детерминистическое приближение,
  которое определяется матрицей сносов.

\begin{figure}%[h]
  \centering
  \includegraphics[width=\linewidth]{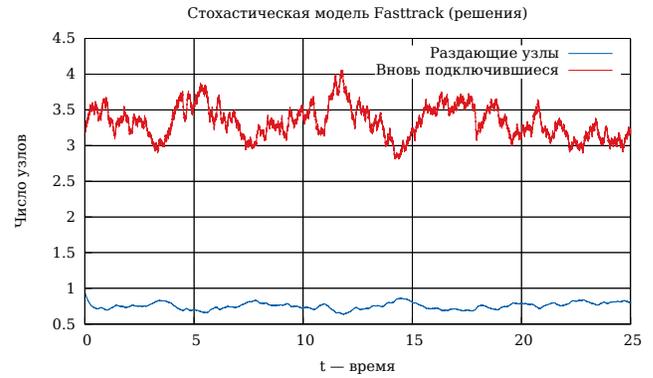}
  \caption{Зависимость числа новых и раздающих узлов от времени в
    сети FastTrack для стохастического случая.}
\label{fig:sft_graph}
\end{figure}

\begin{figure}%[h]
  \centering
  \includegraphics[width=\linewidth]{5}
  \caption{Фазовые портреты стохастической системы Fast Track с
    различными отклонениями $(\Delta x, \Delta y)$ от стационарной
    точки при $\beta \lambda > 4\mu^2$.}
\label{fig:ft5}
\end{figure}

\begin{figure}%[]
  \centering
  \includegraphics[width=\linewidth]{all_sft}
  \caption{Фазовые портреты стохастической системы Fast Track с
    различными отклонениями $(\Delta x, \Delta y)$ от стационарной
    точки при $\beta \lambda > 4\mu^2$.}
\label{fig:all_sft}
\end{figure}

\section{Протокол BitTorrent}

BitTorrent --- пиринговый (P2P) сетевой протокол для кооперативного
обмена файлами через Интернет. Файлы передаются частями, каждый
torrent-клиент, получая (скачивая) эти части, в тоже время отдает
(закачивает) их другим клиентам, что снижает нагрузку и зависимость от
каждого клиента-источника и обеспечивает избыточность данных.

\subsection{Моделирование}

Сначала рассмотрим упрощенную модель закрытой системы, т.е. такую, в
которой не приходят новые клиенты и не уходят раздающие. Кроме того,
сделаем предположение, что файл состоит из одной части. Таким образом
за один шаг взаимодействия нового клиента (личера), желающего скачать
файл, и клиента, раздающего файл (сида), новый клиент скачивает весь
файл и становиться сидом.

Пусть $N$ --- это обозначение нового клиента (личера), $С$ --- это
раздающий клиент (сид), а $\beta$ --- коэффициент
взаимодействия. Тогда схема взаимодействия будет иметь вид:

\begin{equation} 
\label{bt:1}
N+C \xrightarrow{\beta } 2C, \qquad r^{\crd{i}2}=(-1,1).
\end{equation}

Схема отражает, что после взаимодействия личера и сида, в системе
пропадает личер и появляется еще один сид.

Далее, пусть $n$ --- это численность новых клиентов, а $c$ ---
количество сидов в системе.

Запишем вероятности переходов:
\begin{equation} 
\label{bt:2}
s^{+} (n,c) = \beta nc.
\end{equation}

Далее можно записать уравнение Фоккера--Планка для данной модели:
\begin{equation} 
\label{bt:3}
\frac{\partial p(n,c)}{\partial t} = 
{\partial_i} (A^i(n,c) p(n,c))
+\frac{1}{2} {\partial_i \partial_j} (B^{ij}(n,c) p(n,c)),
\end{equation}
где вектор сносов и матрица диффузии имеют следующий вид:
\begin{equation}
  \begin{gathered}
     A^i(n,c)= r^{i\crd{\alpha}}s^+_{\crd{\alpha}} (n,l) ,\\
     B^i(n,c) = r^{i\crd{\alpha}}r^{i\crd{\alpha}} s^+_{\crd{\alpha}} (n,l).
  \end{gathered}
\end{equation}

Таким образом получаем:
\begin{equation} 
\label{bt:4}
  \begin{gathered}
    \mathbf A =
    \begin{pmatrix}
      -1\\
      1
    \end{pmatrix}
    \beta n c +
     =
    \begin{pmatrix}
      - \beta n l\\
      \beta n l 
    \end{pmatrix}, \\
    \begin{multlined}
      \mathbf B =
       \begin{pmatrix}
        -1\\
        1
      \end{pmatrix}
      (-1,1) \beta n c  = 
      \begin{pmatrix}
         \beta n c & - \beta n c \\
        - \beta n c& \beta n c
      \end{pmatrix}.
    \end{multlined}
  \end{gathered}
\end{equation}

Стохастическое дифференциальное уравнение в форме Ланжевена можно
получить воспользовавшись соответствующей формулой.

Также можно записать систему дифференциальных уравнений, описывающую
детерминистическое поведение системы:
\begin{equation} 
\label{bt:5}
\left \{
\begin{aligned}
 \frac{dn}{d t}&=      - \beta n c\\
 \frac{dc}{d t}&=     \beta n c 
\end{aligned}
\right.
\end{equation}

Далее рассмотрим открытую систему, в которой новые клиенты могут
приходить в систему с интенсивностью $\lambda$, а сиды уходить из нее
с интенсивностью $\mu$. Схема взаимодействия имеет вид:

\begin{equation} 
\label{bt:6}
\begin{aligned}
0 \xrightarrow{\lambda } N, & r^{\crd{i}1}=(1,0),\\
N+C \xrightarrow{\beta } 2C, & r^{\crd{i}2}=(-1,1),\\
C  \xrightarrow{\mu } 0, & r^{\crd{i}3}=(0,-1).
\end{aligned}
\end{equation}

Первая строка в схеме описывает появление нового клиента в системе,
вторая строка --- взаимодействие нового клиента и сида, в результате
которого появляется новый сид. А третья --- это уход сида из системы.

Далее, пусть $n$ --- это численность новых клиентов, а $c$ ---
количество сидов в системе.

Эта система с точностью до обозначений совпадает с моделью Fasttrack.

Теперь рассмотрим систему, в которой передаются файлы, состоящие из
$m$ частей. В системе присутствуют следующие участники:

\begin{itemize}
\item Новые клиенты ($N$) --- это клиенты, у которых нет ни одной
  части файла.
\item Личеры ($L$) --- это клиенты, которые уже скачали какое-то
  количество частей файла и могут их раздавать новым клиентам или
  другим личерам.
\item Сиды ($C$) --- это клиенты, у которых есть весь файл, т.е. они
  только раздают.
\end{itemize}

Кроме того $n$ --- это численность новых клиентов, а $c$ ---
количество сидов в системе, $l_i$ --- количество личеров, у которых
есть ровно $i$ частей файла, где $i=\overline{i, n-1}$. Также пусть $
\bar{L}_i$ --- это личеры , у которых есть какие-либо части файла
интересующие личера $L_i$ и соответственно $ \bar{l}_i$i их
количество.

Для данной системы в схеме взаимодействия будут иметь место следующие
типы соотношений:
\begin{equation} 
\label{bt:7}
\begin{aligned}
0  \xrightarrow{\lambda } & N, \\
N+C  \xrightarrow{\beta } & L_1+C, \\
N+L_i  \xrightarrow{\beta_i } & L_1+L_i, \\
L_i + \bar{L}_i  \xrightarrow{\delta_i } & L_{i+1}+\bar{L}_i, \\
L_i + C  \xrightarrow{\gamma_i } & L_{i+1}+C, \\
L_{m-1} + \bar{L}_{m-1}  \xrightarrow{\gamma_{m-1} } & C+\bar{L}_{m-1}, \\
L_{m-1} + C  \xrightarrow{\gamma } & 2C, \\
C  \xrightarrow{\mu } & 0. 
\end{aligned}
\end{equation}

Один шаг взаимодействия --- это передача одной части файла от одного
клиента другому. Первое соотношение описывает появление нового клиента
в системе с интенсивностью $\lambda$. Второе и третье соотношения
описывают взаимодействие нового клиента с сидом или личером с
коэффициентами $\beta$ и $\beta_i$, $(i=\overline{i, m-1})$, в
результате которого новый клиент становиться личером из класса $L_1$
. Четвертое и пятое соотношения --- это взаимодействие личера $L_i$ с
сидом или другим личером с коэффициентами $\delta_i$ и $\gamma_i$
$(i=\overline{i, m-2})$, что приводит к получению личером одной части
файла и переходу его в класс $L_{i+1}$.  Шестое и седьмое описывает
процесс перехода личера в класс сидов с коэффициентами $\gamma_{m-1}$
и $\gamma$, т.е. личер скачивает последнюю часть файл. Последнее
соотношение – это уход сида из системы с интенсивность $\mu$.

Запишем векторы $r^{i\crd{\alpha}}=(n,l_1,l_2,...,l_{m-1},c)$ и 
вероятности перехода $s^+_{\crd{\alpha}}$:

\begin{equation} 
\label{bt:8}
\begin{gathered}
r^{1}  =(1,0,0,...,0), \\
r^{2}  =r_i^3=(-1,1,0,...,0), i=\overline{i, m-1} \\
r_i^4  =r_i^5=(0,...,-1,1,...,0), i=\overline{i, m-2} \\
r^{6}  =r^7=(0,0,...,-1,1), \\
r^{8}  =(0,0,...,-1).
\end{gathered}
\end{equation}
\begin{equation} 
\label{bt:9}
\begin{gathered}
s^{+}_1  =\lambda, \\
s^{+}_2  =\beta n c, \\
s^{+}_{3i}  =\beta_i n l_i, \\
s^{+}_{4i}  =\delta_i l_i \bar{l}_i, i=\overline{i, m-1}\\
s^{+}_{5i}  =\gamma_i l_i c, i=\overline{i, m-2}\\
s^{+}_{6}  =\gamma_{m-1} l_{m-1} \bar{l}_{m-1}, \\
s^{+}_{7}  =\gamma l_{m-1} c, \\
s^{+}_{8}  =\mu c. \\
\end{gathered}
\end{equation}

Для данной модели, аналогично предыдущей, можно записать уравнение
Фоккера-Планка. Но так как детерминистическое поведение полностью
описывается матрицей $A$, запишем только ее.

Таким образом получаем:
\begin{equation}
\label{bt:10}
    \mathbf A =
    \begin{pmatrix}
    \lambda - \beta n c - \sum_{i=1}^{m-1} \beta_i n l_i \\
    \beta n c + \sum_{i=1}^{m-1} \beta_i n l_i -\delta_1 l_1 \bar{l}_1 - \gamma_1 l_1 c \\
      \delta_1 l_1 \bar{l}_1 + \gamma_1 l_1 c -  \delta_2 l_2 \bar{l}_2 - \gamma_2 l_2 c  \\
     \ldots \\
     \begin{multlined}
       \delta_{m-2} l_{m-2} \bar{l}_{m-2} + \gamma_{m-2} l_{m-2} c -
       {} \\ {} -
       \delta_{m-1} l_{m-1} \bar{l}_{m-1} - \gamma_{m-1} l_{m-1} c 
     \end{multlined}
     \\ 
      \delta_{m-1} l_{m-1} \bar{l}_{m-1} + \gamma_{m-1} l_{m-1} c  - \mu c 
    \end{pmatrix}.
\end{equation}

Как следствие можно получить систему дифференциальных уравнений
описывающих динамику численности новых клиентов, личеров и сидов :

\begin{equation} \label{bt:11}
\left \{
\begin{gathered}
  \frac{d n}{d t} =      \lambda - \beta n c - \sum_{i=1}^{m-1} \beta_i n l_i, \\
 \frac{d l_1}{d t}=     \beta n c + \sum_{i=1}^{m-1} \beta_i n l_i -\delta_1 l_1 \bar{l}_1 - \gamma_1 l_1 c, \\ 
 \frac{d l_2}{d t}=  \delta_1 l_1 \bar{l}_1 + \gamma_1 l_1 c -  \delta_2 l_2 \bar{l}_2 - \gamma_2 l_2 c,  \\
    \ldots \\
\begin{multlined}
 \frac{d l_{m-1}}{d t}=  \delta_{m-2} l_{m-2} \bar{l}_{m-2} + 
 \gamma_{m-2} l_{m-2} c - {} \\ {} - 
 \delta_{m-1} l_{m-1} \bar{l}_{m-1} -
 \gamma_{m-1} l_{m-1} c,
\end{multlined}
\\ 
\frac{d c}{d t}= \delta_{m-1} l_{m-1} \bar{l}_{m-1} + \gamma_{m-1} l_{m-1} c  - \mu c. 
\end{gathered}
\right.
\end{equation}

Сделаем предположение, что
$\delta=\delta_{1}=\delta_{2}=...=\delta_{m-1}=const$. Сложим в
системе уравнения со второго по $m+1$ и при обозначении всех личеров и
сидов через $l = l_1 + l_2 + ... + l_{m-1} + c$ получим упрощённую
систему следующего вида:

\begin{equation} 
\label{bt:12}
\left \{
\begin{aligned}
 \frac{d n}{d t}&=      \lambda - \beta n (l+c), \\
 \frac{d (l+c)}{d t}&=     \beta n (l+c)  - \mu c. 
\end{aligned}
\right.
\end{equation}

\section{Заключение}

\begin{enumerate}
\item В работе описан метод получения стохастических моделей для
  систем, которые возможно описывать одношаговыми
  процессами. Предложенный метод позволяет получить универсальные
  правила записи стохастических дифференциальных уравнений для систем,
  процессы в которых представимы как одношаговые процессы. А также
  расширить аппарат инструментов, используемых для анализа модели, так
  как одновременно при применении данного подхода для описания системы
  можно получить обыкновенное стохастическое дифференциальное
  уравнение и уравнение в частных производных в форме уравнения
  Фоккера--Планка.
\item Изучено влияния введения стохастики в детерминистические модели,
  на примере модели протокола FastTrack и Bittorrent. Полученные
  результаты показывают, что введение стохастики в стационарном режиме
  слабо влияет на поведение системы, поэтому при ее изучении можно
  рассматривать детерминистическую модель. Кроме того, как показал
  рассмотренный пример в некоторых случаях для изучения системы можно
  рассматривать ее детерминистическое приближение, которое
  определяется матрицей сносов.
\end{enumerate}

\bibliographystyle{gost2008l}

\bibliography{bib/p2p_sdu/p2p_sdu}